\documentclass[preprint,nofootinbib,showpacs]{revtex4}
\usepackage{natbib}
\usepackage{times}
\usepackage{amssymb,amsbsy,amsmath,amsfonts}
\usepackage{graphicx}
\usepackage{float}
\usepackage{morefloats}
\usepackage{rotating}
\usepackage{srcltx}
\usepackage{slashed}
\begin{document}

\title{SU(3) breaking corrections to the $D$, $D^*$, $B$, and  $B^*$ decay constants }

\author{ M. Altenbuchinger$^1$, L.S. Geng,$^{2,1}$   and W. Weise$^1$}
\affiliation{
$^1$Physik Department, Technische Universit\" at M\"unchen, D-85747 Garching, Germany\\
$^2$School of Physics and Nuclear Energy Engineering, Beihang University,  Beijing 100191,  China}

 \begin{abstract}
We report on a first next-to-next-to-leading order calculation of the decay constants of the $D$ ($D^*$) and $B$ ($B^*$) mesons using a covariant formulation of chiral perturbation theory. It is shown that, using the state-of-the-art lattice QCD results on $f_{D_s}/f_D$ as input, one can predict
quantitatively the ratios of $f_{D_s^*}/f_{D^*}$, $f_{B_s}/f_B$, and $f_{B^*_s}/f_{B^*}$ taking into account heavy-quark spin-flavor symmetry breaking effects on the relevant low-energy constants. The predicted relations between these ratios, $f_{D^*_s}/f_{D^*}<f_{D_s}/f_D$ and $f_{B_s}/f_B>f_{D_s}/f_D$,  and their light-quark mass dependence should be testable in future lattice QCD simulations,  providing
a stringent test of our understanding of heavy quark spin-flavor symmetry, chiral symmetry and their breaking patterns. \end{abstract}


\date{\today}

\maketitle

The decay constants of the ground-state $D$ ($D^*$) and $B$ ($B^*$) mesons have been subjects of intensive study over the past two decades. Assuming exact isospin symmetry, there  are eight independent heavy-light (HL) decay constants: $f_D$ ($f_{D^*}$), $f_{D_s}$ ($f_{D^*_s}$), $f_B$ ($f_{B^*}$), $f_{B_s}$ ($f_{B^*_s}$).
In the static limit of infinitely heavy charm (bottom) quarks, the vector and pseudoscalar D (B) meson decay constants become degenerate, and in the chiral limit of massless up, down and strange quarks, the strange and non-strange D (B) meson decay constants become degenerate. In the real world, both limits are only approximately realized and, as a result, the degeneracy disappears.

The gluonic sector of Quantum ChromoDynamics (QCD) is flavor blind, so the non-degeneracy between the HL decay constants must be entirely due to finite values of the quark masses in their hierarchy. A systematic way of studying the effects of finite quark masses is the heavy-meson chiral perturbation theory (HM ChPT)~\cite{Wise:1992hn,Yan:1992gz,Burdman:1992gh}. The HL decay constants have been calculated up to next-to-leading order (NLO) in the chiral expansion, and to leading-order (LO)~\cite{Grinstein:1992qt,Goity:1992tp} and NLO~\cite{Grinstein:1993ys,Boyd:1994pa} in $1/m_H$ expansion, where $m_H$ is the generic mass of the HL systems.  In a recent work, a covariant formulation of ChPT has been employed to study the pseudoscalar decay constants up to NNLO for the first time and faster convergence compared to HM ChPT was observed~\cite{Geng:2010df}.

Lattice QCD (LQCD) provides an ab initio method for calculating the HL decay constants. There exist many $n_f=2+1$ computations of the pseudoscalar decay constants, $f_{D_s}$ and $f_D$~\cite{Aubin:2005ar,Follana:2007uv,Davies:2010ip,Simone:2010zz,Namekawa:2011wt}, and $f_{B_s}$ and $f_B$~\cite{Gamiz:2009ku,Simone:2010zz,Albertus:2010nm}, motivated by the important role they play  in determinations of the CKM matrix elements and in tests of the standard model (see, e.g., Ref.~\cite{Dobrescu:2008er}).
 On the other hand, for the vector meson decay constants, most existing simulations are quenched~\cite{Becirevic:1998ua,Bowler:2000xw,Bernard:2001fz}, except for Ref.~\cite{Collins:1999ff} where $n_f=2$. Simulations with $n_f=2+1$ are underway~\cite{Davies:2010}.

 In this letter, we report on a first next-to-next-to-leading order (NNLO) covariant ChPT study of the HL pseudoscalar and vector meson decay constants. We will show that heavy-quark spin-flavor symmetry breaking effects only lead to small deviations of  the ratios $f_{B_s}/f_B$, $f_{D^*_s}/f_{D^*}$, and $f_{B^*_s}/f_{B^*}$, from $f_{D_s}/f_D$. Utilizing the latest HPQCD data on $f_{D_s}$ and $f_D$~\cite{Follana:2007uv}, and taking into account heavy-quark spin-flavor symmetry breaking corrections to the relevant low-energy constants (LECs), we are able to make some highly nontrivial predictions on the other three ratios. The predicted light-quark mass dependencies of  the HL decay constants  are also of great value for future lattice simulations.

The decay constants of the $D$ and $D^*$ mesons with quark content $\bar{q}c$, with $q=u,d,s$, are defined as
\begin{eqnarray}
 \langle 0|\bar{q}\gamma^\mu\gamma_5 c(0)|P_q(p)\rangle&=&-i f_{P_q} p^\mu,\\
  \langle0|\bar{q}\gamma^\mu c(0)|P^*_q(p,\epsilon)\rangle&=& F_{P^*_q} \epsilon^\mu,
\end{eqnarray}
where $P_q$ denotes a pseudoscalar meson and $P^*_q$ a vector meson. In this convention, $f_{P_q}$  has mass dimension one and $F_{P^*_q}$ has mass dimension two~\cite{Manohar:2000dt}.  For the sake of comparison with other approaches, we introduce $f_{P^*}=F_{P^*}/m_{P^*}$, which has mass dimension one.  Our formalism can be trivially extended to the $B$ meson decay constants and therefore in the following we concentrate on the $D$ mesons.

To construct the relevant Lagrangians in a compact manner, one introduces the following fields\footnote{It should be noted that
the heavy-light states in the relativistic formalism have mass dimension of 1 instead of $3/2$ as in the HM formulation.} and currents as in Ref.~\cite{Burdman:1992gh}:
\begin{eqnarray}
H&=&\frac{i\slashed{d} + m_P}{2 m_P} (\gamma^\mu P^*_\mu+i P\gamma^5),\\
J&=&\frac{1}{2}\gamma^\mu(1-\gamma_5)J_\mu,
\end{eqnarray}
where $P=(D^0,D^+,D^+_s)$, $P^*_\mu=(D^{*0},D^{*+},D^{*+}_s)$,  $J_\mu=(J^{uc}_\mu,J^{dc}_\mu,J^{sc}_\mu)^T$ with the weak current $J^{qc}_\mu=\bar{q}\gamma_\mu(1-\gamma^5)c$, $m_P$ is the characteristic mass of the $P$ triplet introduced to conserve heavy quark spin-flavor symmetry in the $m_P\rightarrow\infty$ limit: $\mathring{m}_D$ at NLO and $m_D$ at NNLO (see Table 1). The covariant derivative is
defined as $d_\mu=\partial_\mu + \Gamma_\mu$ with $\Gamma_\mu=\frac{1}{2}(u^\dagger\partial_\mu u+ u\partial_\mu u^\dagger$) and $u^2=U=\exp[\frac{i\Phi}{F_0}]$ with $\Phi$ the pseudoscalar octet matrix of Nambu-Goldstone (NG) boson fields, $F_0$ their decay constant in the chiral limit.
 The weak couplings have the following form~\cite{Burdman:1992gh}:
\begin{eqnarray}\label{eq:lhc}
 \mathcal{L}^{(1)}_w&=&\alpha  \mathrm{Tr}[ J_b H_a] u^\dagger_{ab},\\
\mathcal{L}^{(2)}_w&=& \frac{\alpha}{\Lambda_\chi}\left\{i\beta_1\mathrm{Tr}[J_b H_a \slashed{\omega}_{ab}]
+\frac{\beta_2}{m_P}\mathrm{Tr}[J_b \partial_\nu H_a] \omega^\nu_{ab}\right\},\\
 \mathcal{L}^{(3)}_w&=&-\frac{\alpha}{2\Lambda_\chi^2}\left\{b_D\mathrm{Tr}[J_b H_a](\chi_+ u^\dagger)_{ab}\right.\nonumber\\
 &&\left.\hspace{2cm}+b_A
\mathrm{Tr}[J_b H_a] u^\dagger_{ab} (\chi_+)_{cc}\right\},
\end{eqnarray}
where $\alpha$ is a normalization constant of mass dimension two,
 $\omega_\mu=u\partial_\mu U^\dagger$,
   $\Lambda_\chi=4\pi F_0$ is the scale of spontaneous chiral symmetry breaking,
   and
   $\chi_+=u^\dagger \chi^\dagger u^\dagger+ u\chi u$ with $\chi=\mathcal{M}=\mathrm{diag}(m_\pi^2,m_\pi^2,2 m_K^2-m_\pi^2)$. Here
and in the following  $\mathrm{Tr}$ denotes trace for the Dirac matrices.
In Eqs.~(5,6,7), the superscript in $\mathcal{L}$ denotes the chiral order of the corresponding Lagrangian. Here we have counted the axial current, the derivative on the NG boson fields, and their masses as
$\mathcal{O}(p)$, as usual. 
\begin{figure}[t]
\centerline{\includegraphics[scale=0.7]{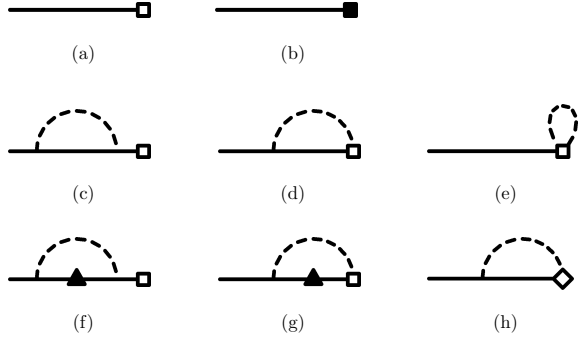}}
\caption{Feynman diagrams contributing to the heavy-light (HL) decay constants up to NNLO: (a) and (b) are LO and NLO tree level
diagrams, loop diagrams  (c), (d) and (e) contribute at NLO while diagrams
(f), (g) and (h) contribute at NNLO. The solid lines denote either HL pseudoscalar or HL vector mesons and combinations thereof,
dashed lines represent Nambu-Goldstone bosons, the empty (solid) squares and empty diamond  denote current from the first (third) and second order Lagrangians, and the solid triangles denote mass insertions of second chiral order (see Ref.~\cite{Geng:2010df}).}\label{fig:feyn}
\end{figure}

To calculate chiral loops, the following LO Lagrangian is introduced~\cite{Wise:1992hn,Yan:1992gz,Burdman:1992gh,Geng:2010vw} (in this letter, only the relevant terms are explicitly shown):
\begin{equation}\label{eq:gddstarpi}
 \mathcal{L}^{(1)}=\frac{g m_P}{2}\mathrm{Tr}[\bar{H}_b H_a \slashed{u}_{ab} \gamma_5].
\end{equation}
It describes the interactions between a pair of HL mesons ($PP^*$ or $P^*P^*$) with a Nambu-Goldstone boson $\phi=\pi, K,\eta$. In Eq.~(\ref{eq:gddstarpi}),
we have introduced $m_P$ for the sake of convenience. It should be taken as $\mathring{m}_D$ ($\mathring{m_B}$) at NLO and $m_D$ ($m_B$) at NNLO.
In the $D$ meson sector,  $g_{DD^*\pi}\equiv g=0.60\pm0.07$~\cite{Geng:2010vw}, while
$g_{D^*D^*\pi}\equiv g^*$ is not precisely known.
At the chiral order we are working, one can take $g_{DD^*\phi}=g_{DD^*\pi}$. If heavy quark spin-flavor symmetry is exact, $g_{BB^*\phi}=g_{B^*B^*\phi}=g_{D^*D^*\phi}=g_{DD^*\phi}$, otherwise deviations are expected.

The Feynman diagrams contributing to the decay constants
up to NNLO \footnote{The chiral order of a properly renormalized diagram with $L$ loops, $N_M$ ($N_H$) Nambu-Goldstone boson (HL meson) propagators and $V_k$ vertices from $k$th-order Lagrangians is $n_{\chi PT}=4L-2N_M-N_H+\sum_k kV_k$.}are shown in Fig.~\ref{fig:feyn}.
For the HL pseudoscalar meson decay constants, diagrams (a-g) have been calculated in Ref.~\cite{Geng:2010df}. However, diagram (h) that contains two new LECs
$\beta_1$ and $\beta_2$ was not considered there. Its contribution to the pseudoscalar decay constant is
\begin{equation}
 R^h_i=\frac{\alpha}{\Lambda_\chi}\sum\limits_{j,k}\xi_{i,j,k} \left(\frac{g m_P }{16 F_0^2 m_i^2} \right)\left(\frac{-1 }{16\pi^2}\right) \phi^h(m_i^2,m_k^2)\nonumber
\end{equation}
with
\begin{eqnarray}
\phi^h&=&
4\beta_1\Big[m_k^2 ((4
   m_i^2-m_k^2)\bar{B}_0(m_i^2,m_i^2,m_k^2)+\bar{A}_0(m_k^2))\nonumber\\
   &&+(2 m_i^2-m_k^2)
   \bar{A}_0(m_i^2)\Big]+\frac{\beta_2}{m_i^2}\Big[
   -2 m_k^4 (m_k^2-4
   m_i^2)\nonumber\\
   &&\times
   \bar{B}_0(m_i^2,m_i^2,m_k^2)-m_i^6+\left(4 m_i^2
   m_k^2+6 m_i^4-2 m_k^4\right)\nonumber\\
   &&\times \bar{A}_0(m_i^2)+2 (5 m_i^2
   m_k^2+m_k^4) \bar{A}_0(m_k^2)+m_i^2
   m_k^4\Big],\nonumber
 \end{eqnarray}
where $\xi_{i,j,k}$ can be found in Table 2 of Ref.~\cite{Geng:2010df} with $i$ running over $D$ and $D_s$, $j$ over
$D^*$ and $D^*_s$, and $k$ over $\pi$, $\eta$, and $K$. The functions $\bar{A}_0=(-16\pi^2)A_0$ and
$\bar{B}_0=(-16\pi^2)B_0$ with $A_0$ and $B_0$ defined in the appendix of Ref.~\cite{Geng:2010df}. It should be noted that at NNLO the HL meson masses appearing here are  the average of the vector and pseudoscalar HL mesons, i.e. $\mathring{m}_D$ and $\mathring{m}_B$ in Table 1. For the diagrams contributing to the HL vector meson decay constants, the computation of the corresponding diagrams (a, b, e) is the same as in the case of the pseudoscalar decay constants, keeping in mind that now $\alpha$ , $b_D$, and $b_A$ are all understood to be different from those in the pseudoscalar sector by heavy-quark spin symmetry breaking corrections.
\begin{table*}[t]
      \renewcommand{\arraystretch}{1.6}
     \setlength{\tabcolsep}{0.2cm}
     \centering
     \caption{\label{table:par}Numerical values of the isospin-averaged masses~\cite{pdg2010} and decay constants (in units of MeV) used in the present study.
 The eta meson mass is calculated using the Gell-Mann-Okubo mass relation: $m_\eta^2=(4 m_K^2-m_\pi^2)/3$. $F_0$ is
 the average of physical $f_\pi$, $f_K$ and $f_\eta$.}
     \begin{tabular}{ccccccccccccc}
     \hline\hline
    $\mathring{m}_{D}$     & $m_D$    &$\Delta_s$ &   $\Delta$  & $\mathring{m}_B$ & $m_B$ & $\Delta_s(B)$ & $\Delta(B)$ &  $m_\pi$ & $m_K$  &    $m_\eta$   &   $f_\pi$ &  $F_0$  \\
   1972.1     & 1867.2         &  102.5   & 142.6 &
   5331.8 & 5279.3 & 88.7 & 47.5 & 138.0  & 495.6 & 566.7 & 92.4 & $1.15 f_\pi$\\

 \hline\hline
    \end{tabular} 
\end{table*}

The loop diagrams for vector mesons fall into two categories, depending on whether a HL vector meson (class I) or a HL
pseudoscalar meson (class II) propagates in the loop. For vector mesons,
the wave function renormalization diagrams (f)  yield:
\begin{eqnarray}
 R^{f^{I,II}}_i&=&\sum_{j,k}\xi_{i,j,k}\left(\frac{1 }{18 F_0^2}\right)\left(\frac{-1 }{16\pi^2}\right)\frac{d\,\phi^{f^{I,II}}(p_i^2,m_j^2,m_k^2)}{d\,p_i^2}\Big|_{p_i^2=m_i^2},\nonumber
\end{eqnarray}
with
\begin{eqnarray}
 \phi^{f^I}&=&(g^*)^2\Bigg[3 (-p_i^2+(m_j-m_k)^2) (-p_i^2+(m_j+m_k)^2)\nonumber\\
   &&\times \bar{B}_0(p_i^2,m_k^2,m_j^2)+3
   \bar{A}_0(m_j^2)(-p_i^2+m_k^2-m_j^2)\nonumber\\
   &&-3
   \bar{A}_0(m_k^2)(p_i^2+m_k^2-m_j^2)   +p_i^2 (-p_i^2+3
   m_k^2+3 m_j^2)\Bigg],\nonumber
   \end{eqnarray}
   \begin{eqnarray}
\phi^{f^{II}}&=&-\frac{m_P^2}{2p_i^2}g^2\Big[-3 (-2 m_k^2
   (p_i^2+m_j^2)+(m_j^2-p_i^2){}^2
   +m_k^4)\nonumber\\
&&\times \bar{B}_0(p_i^2,m_k^2,m_j^2)  +3
   \bar{A}_0(m_k^2)
   (p_i^2+m_k^2-m_j^2)\nonumber\\
   &&+3
   \bar{A}_0(m_j^2)(p_i^2-m_k^2+m_j^2)
   +6 p_i^2
   (m_k^2+m_j^2)-2 p_i^4\Big],\nonumber
   \end{eqnarray}
where  $i$ denotes $(D^*,D^*_s)$ and $j$ denotes either $(D^*,D_s^*)$ or $(D,D_s)$.

Diagrams (g) yields $R^{g^{I}}=0$ and
\begin{equation}
 R^{g^{II}}=\sum\limits_{j,k}\xi_{i,j,k}\left(\frac{\alpha  g}{72 F_0^2 m_i^2}\right)\left(\frac{-1 }{16\pi^2}\right)\phi^{g^{II}}(m_i^2,m_j^2,m_k^2)\nonumber
\end{equation}
with
\begin{eqnarray}
\phi^{g^{II}}&=& -3 ((m_i-m_k)^2-m_j^2)
    ((m_i+m_k)^2-m_j^2)\nonumber\\
    &&\times \bar{B}_0(m_i^2,m_j^2,m_k^2)+3
   \bar{A}_0(m_j^2)
 (m_i^2+m_j^2-m_k^2)\nonumber\\
 &&+3 \bar{A}_0(m_k^2)
   (m_i^2-m_j^2+m_k^2)-2 m_i^2
   (m_i^2-3(m_j^2+m_k^2)).\nonumber
\end{eqnarray}

Diagrams (h) give
\begin{eqnarray}
 R^{h^{I,II}}&=&\frac{\alpha}{\Lambda_\chi}\sum\limits_{j,k}\xi_{i,j,k}\left(\frac{g m_P}{144 F_0^2 m_i^2}\right)\left(\frac{-1 }{16\pi^2}\right)
\phi^{h^{I,II}}(m_i^2,m_k^2)\nonumber
\end{eqnarray}
with
\begin{eqnarray}
 \phi^{h^I}&=&8\beta_1\frac{m_i^2}{m_P^2}\frac{g^*}{g}\Big[(6 m_i^2-3 m_k^2)
   \bar{A}_0(m_i^2)-3 m_i^2 m_k^2-2
   m_i^4   \nonumber\\
   && +3 m_k^2 \big[(4
   m_i^2-m_k^2)
   \bar{B}_0(m_i^2,m_i^2,m_k^2)+ \bar{A}_0(m_k
   ^2)\big]\Big],\nonumber
   \end{eqnarray}
   \begin{eqnarray}
 \phi^{h^{II}}&=&
4\beta_1\Big[3 m_k^2
   ((4 m_i^2-m_k^2)
   \bar{B}_0(m_i^2,m_i^2,m_k^2)\nonumber\\
   &&+\bar{A}_0(m_k
   ^2))+(6 m_i^2-3 m_k^2)
   \bar{A}_0(m_i^2)+6 m_i^2 m_k^2\nonumber\\
   &&+4
   m_i^4\Big]+\frac{\beta_2}{m_P^2} \Big[-6 m_k^4 (m_k^2-4
   m_i^2) \bar{B}_0(m_i^2,m_i^2,m_k^2)\nonumber\\
   &&+6
   (3 m_i^2-m_k^2) (m_i^2+m_k^2)
   \bar{A}_0(m_i^2)+8
   m_i^4 m_k^2\nonumber\\
   &&+21 m_i^2 m_k^4+9 m_i^6
   +6 (3 m_i^2
   m_k^2+m_k^4) \bar{A}_0(m_k^2)\Big].\nonumber
   \end{eqnarray}

As explained in Ref.~\cite{Geng:2010df}, mass insertions
in diagrams (c, d)  generate
NNLO contributions. Therefore, using
 $m_{D_s}\rightarrow m_D+\Delta_s$, $m_{D^*}\rightarrow m_D+\Delta$, and
  $m_{D^*_s}\rightarrow m_D+\Delta+\Delta_s$
for the HL meson masses in diagrams (f, g), one obtains the full NNLO results of these diagrams. The complete NNLO results for the pseudoscalar and vector HL decay constants are
\begin{eqnarray*}
 f_i&=&\hat{\alpha}(1+\tilde{Z}_i/2)+\delta_i+T_i+\tilde{C}_i+\tilde{R}^h_i,\\
 F^*_i&=&\alpha(1+(\tilde{R}^{f^I}_i+\tilde{R}^{f^{II}}_i)/2)+\delta_i+\tilde{R}^{g^{II}}_i+R^e_i+\tilde{R}^{h^I}_i+\tilde{R}^{h^{II}}_i,
\end{eqnarray*}
where $\hat{\alpha}=\alpha/m_P$ and $Z_i$, $T_i$, and $C_i$ can be found in Ref.~\cite{Geng:2010df}. The ``tilde'' indicates that one has to perform a subtraction to remove the power-counting-breaking terms that are inherent of covariant ChPT involving heavy hadrons whose masses do not vanish at the chiral limit (for details see Refs.~\cite{Geng:2010vw,Geng:2010df}). Furthermore, a second subtraction  is needed to ensure that heavy-quark spin-flavor symmetry is exact in the limit of infinitely heavy quark masses. Details and consequences for phenomenology will be reported in a separate work.
After these subtractions the results can be expanded in the inverse heavy-light meson mass. In the limit $m_P\rightarrow \infty$ the lowest order HMChPT results are recovered. The covariant approach, being fully relativistic, sums all powers of contributions in $1/m_P$, which are of higher order in HMChPT. Such a relativistic formulation is not only formally appealing. It also converges faster than  non-relativistic formulations, such as HMChPT and HBChPT. This has been recently demonstrated in the one-baryon sector and in heavy-light systems for a number of observables (see, e.g., Refs.~\cite{Geng:2010vw,Geng:2010df}.
It should be stressed that the loop functions are divergent and the infinities have been removed by the standard
$\overline{\text{MS}}$ procedure, as in Ref.~\cite{Geng:2010df}.

\begin{figure}[b]
\centerline{\includegraphics[scale=0.7]{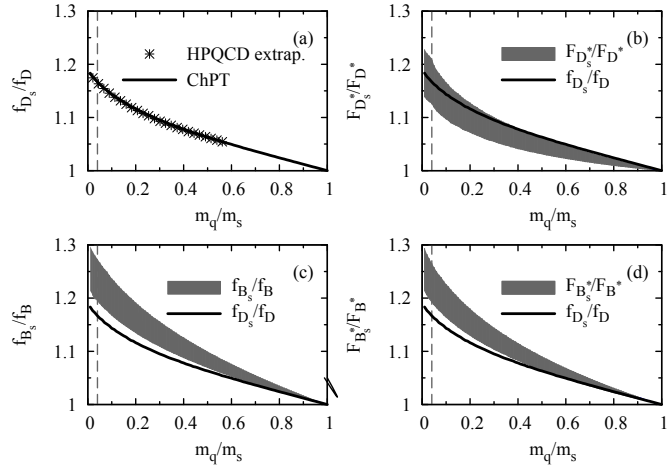}}
\caption{Light-quark mass evolution of $f_{D_s}/f_D$, $F_{D^*_s}/F_{D^*}$,
$f_{B_s}/f_B$, and $F_{B^*_s}/F_{B^*}$. The ratio $r=m_q/m_s$ is related to the pseudoscalar meson masses at leading chiral order through
$m_\pi^2=2B_0 m_s r$ and $m_K^2=B_0 m_s (r+1)$ with $B_0=m_\pi^2/(2m_q)$, where $m_s$ is  the
physical strange quark mass and $m_q$ the average of up and down quark masses. The vertical dotted lines denote
physical $m_q/m_s$.
 \label{fig:res}}
\end{figure}

Now we are in a position to perform numerical studies.
We first fix the five LECs, $\alpha$, $b_D$, $b_A$, $\beta_1$, and $\beta_2$, by fitting the HPQCD $f_{D_s}/f_D$ extrapolations~\cite{Follana:2007uv}. The results are shown in Fig.~(\ref{fig:res}a). The  NNLO ChPT fits the chiral and continuum extrapolated lattice QCD results remarkably well, keeping in mind that the HPQCD extrapolations were obtained using the NLO HMChPT results supplemented with higher-order analytical terms~\cite{Follana:2007uv}.

In addition to providing the NNLO ChPT results that should be useful for future
lattice simulations of the HL decay constants, a primary aim of the present study is to predict quantitatively
the SU(3) breaking corrections to $f_{D^*_s}/f_{D^*}$, $f_{B_s}/f_B$, and $f_{B^*_s}/f_{B^*}$ from that of the $f_{D_s}/f_D$. To achieve this, one must take into account heavy-quark spin-flavor symmetry breaking corrections to the LECs: $\alpha$, $b_D$, $b_A$, $\beta_1$, $\beta_2$, and $g_{PP^*\phi}$ ($g_{P^*P^*\phi}$).

 The LEC $\alpha$ is only relevant for the absolute value of the decay constants, therefore it does not appear  in the SU(3) breaking ratios. However, in the Lagrangian of Eqs.~(5,6), one implicitly assumes heavy-quark spin symmetry, i.e., $c'=\frac{f_{P^*} \sqrt{m_{P^*}}}{f_P \sqrt{m_P}}=1$, which affects the computation of loop diagrams (g) for pseudoscalars and (g, h) for vector mesons (see Ref.~\cite{Geng:2010df} for details). Recent quenched LQCD simulations suggest that $c'$ is within the range of $1.0\sim1.2$  ~\cite{Becirevic:1998ua,Bowler:2000xw}.
To be conservative we allow $c'$ to vary within $0.8\sim1.2$.  For $b_D$, $b_A$, $\beta_1$, and $\beta_2$, no LQCD data are available. However, the corrections to those constants from heavy-quark spin-flavor symmetry breaking are expected to be $\lesssim20\%$.

The LECs that affect  the predicted ratios  most prominently turn out to be $g$ and $g^*$, which determine the size of chiral loop contributions. In the present case $g_{DD^*\pi}$ is determined  by reproducing the $D^*$ meson decay width.
Recent $n_f=2$ LQCD simulations suggest that $g_{BB^*\pi}$ is  in the range of
$0.4\sim0.6$ ~\cite{Ohki:2008py,Becirevic:2009yb,Bulava:2010ej}. We therefore take the central value of $0.516$ from Ref.~\cite{Ohki:2008py} and assign a 20\% uncertainty.  Studies based on QCD sum rules indicate that $g$ and $g^*$ could differ by $10\sim20\%$~\cite{DiBartolomeo:1994ir,Bracco:2011pg}. We take this into account in our study.

With heavy-quark spin-flavor symmetry breaking effects on the relevant LECs taken into account as described above, we can now make predictions for the ratios of $f_{B_s}/f_B$, $F_{D^*_s}/F_{D^*}$, and $F_{B^*_s}/F_{B^*}$ and their light-quark mass dependencies. The results are shown in Figs.~(\ref{fig:res}b,\ref{fig:res}c,\ref{fig:res}d). The differences between the four ratios are small, at the order of a few percent. Interestingly, the ratios of the B meson decay constants are found to be larger than those of their D counterparts, in agreement with the HPQCD results~\cite{Follana:2007uv,Gamiz:2009ku}. Fully dynamical lattice simulations of the vector meson decay constants should provide a stringent test of our predictions. It should be stressed that the bands shown in Fig.~\ref{fig:res} reflect the estimated effects of heavy-quark spin-flavor symmetry breaking from the change of the relevant LECs, in addition to those induced by the covariant formulation of ChPT, the use of physical mass splittings and  different $g_{DD^*\phi}$ ($g_{BB^*\phi}$). The same is true for the uncertainties of our results given in Table II.

\begin{table}[t]
\caption{Ratios of $f_{D_s}/f_D$, $f_{D^*_s}/f_{D^*}$, $f_{B_s}/f_B$, and $f_{B^*_s}/f_{B^*}$ from different approaches.
The $f_{D_s}/f_D=1.164$ from the HPQCD collaboration~\cite{Follana:2007uv} is used as input in our approach.\label{table:res}}
\begin{center}
\begin{tabular}{lllllll}
\hline\hline
Ref. &$f_{D_s}/f_D$&$f_{D^*_s}/f_{D^*}$&$f_{B_s}/f_B$& $f_{B^*_s}/f_{B^*}$\\
\hline
PDG~\cite{pdg2010}&$1.25(6)$&-&-&-\\
FCM~\cite{Badalian:2007km}&$1.24(4)$&$1.12$&$1.19(3)$&$1.15$\\
RQM~\cite{Ebert:2006hj}&$1.15$&$1.02$&$1.15$&$1.15$\\
LFQM~\cite{Choi:2007se}& $1.18(1.20)$ & $1.14(1.18)$ & $1.24(1.32)$ & $1.23(1.32)$ \\
QLQCD~\cite{Becirevic:1998ua}&$1.10(2)$&$1.11(3)$&$1.14(3)(1)$&$1.17(4)(3)$\\
QLQCD~\cite{Bowler:2000xw}&$1.11(1)(1)$&$1.09(1)(2)$&$1.13(1)(1)$&$1.14(2)(2)$\\
LQCD~\cite{Collins:1999ff}&  & & $1.14(2)(2)$ & $1.14(2)(2)$\\
HPQCD~\cite{Follana:2007uv,Gamiz:2009ku}&$1.164(11)$&&$1.226(26)$&\\
NNLO ChPT&$1.17$&$1.10(5)$&$1.24(4)$&$1.20(4)$\\
\hline\hline
\end {tabular}
\end{center}
\end{table}
Our predicted ratios at the physical point are compared  in Table II with the results from a number of other approaches, including the lattice simulations~\cite{Becirevic:1998ua,Bowler:2000xw,Collins:1999ff}, the relativistic quark model (RQM)~\cite{Ebert:2006hj}, the light-front quark model (LFQM)~\cite{Choi:2007se}, and the field correlator method (FCM)~\cite{Badalian:2007km}.\footnote{It should be mentioned that
the NNLO ChPT predictions cover the NLO predictions within uncertainties.} Our predictions for the relative magnitude of the $f_{P^*_s}/f_{P^*}$ vs. $f_{P_s}/f_P$ ratios agree with those
of the FCM~\cite{Badalian:2007km}, the RQM~\cite{Ebert:2006hj} and LFQM~\cite{Choi:2007se}. It should be noted that the results in Fig.~\ref{fig:res} are obtained with a renormalization scale of 1 GeV~\cite{Geng:2010df}. Uncertainties have been estimated changing this scale between $\mu=m_D$ and $\mu=m_B$ for the calculation of  $D$ and $B$ decay constants, respectively. The changes turn out to be small and are taken into account in the results shown in Table II.

In summary, we have calculated the pseudoscalar and vector decay constants of the $B$ and $D$ mesons using a covariant formulation of chiral perturbation theory up to next-to-next-to-leading order and found that it can describe well the HPQCD $n_f=2+1$ data on $f_{D_s}/f_D$
. Taking into account heavy-quark spin-flavor symmetry breaking effects on the relevant LECs,
we have made predictions for the ratios of $f_{B_s}/f_B$, $f_{D^*_s}/f_{D^*}$, and $f_{B^*_s}/f_{B^*}$ and their light quark mass dependencies that should be testable in the near future. Our results show that  $f_{B_s}/f_B>f_{D_s}/f_D$ and $f_{D^*_s}/f_{D^*}<f_{D_s}/f_D$ in a large portion of the allowed parameter space.

This work is supported in part by BMBF, the A.v. Humboldt foundation, the Fundamental Research Funds for the Central Universities, the National Natural
Science Foundation of China (Grant No. 11005007), and by the DFG Excellence Cluster ``Origin and Structure of the Universe."

\end{document}